\documentclass[twocolumn,showpacs,showkeys,preprintnumbers,amsmath,amssymb,prb]{revtex4}
\usepackage[svgnames]{xcolor}
\usepackage{graphicx, tikz, pgfplots}
\usepackage{graphicx}
\usepackage{dcolumn}
\usepackage{bm}

\usepackage[colorlinks=true, linkcolor=blue, citecolor=blue, hyperindex=true, linktocpage=true]{hyperref}

\begin{document}
\preprint{APS/123-QED}

\title{Undergraduate experiment in superconductor point-contact spectroscopy with a Nb/Au junction }
\author{Lucas Janson, Matthew Klein, Heather Lewis, Andrew Lucas, Andrew Marantan, Katherine Luna}
\email{kluna@stanford.edu}
\affiliation{Department of Physics, Stanford University, Stanford CA 94305-4045, USA} 
\date{\today}

\begin{abstract}
We describe an experiment in superconductivity suitable for an advanced undergraduate laboratory.  Point-contact spectroscopy is performed by measuring the differential conductance between an electrochemically etched gold tip and a 100-nm superconducting niobium film with a transition temperature $T_\mathrm{c}\approx$ 7 K.  By fitting the results to Blonder-Tinkham-Klapwijk theory using a finite lifetime of quasiparticles,  we obtain a superconducting gap energy $\Delta \approx 1.53$ meV, a lower bound to the Fermi velocity $v_\mathrm{F} \ge 3.1 \times 10^7$ cm/s, and a BCS coherence length $\xi \approx 43$ nm for niobium.  These results are in good agreement with previous measurements.
\end{abstract}

\maketitle

The following article has been accepted by the American Journal of Physics. After it is published, it will be found at http://scitation.aip.org/ajp/.

\section{Introduction}

The resistivity of a normal metal (N) on distance scales greater than the mean free path $\ell$ is characterized by Drude theory.  Because the Drude model is
based on the diffusive motion of electrons, the resistance scales with material or probe geometry. On length scales shorter than $\ell$, electrons move ballistically and the resistance is not so simply related to the geometry. This ballistic regime, which is becoming increasingly relevant with the miniaturization of electronic devices, can be accessed by tunneling into a material from a contact of size $a \ll \ell$. In such contacts, rich spectroscopic information may be obtained from measurements of differential conductance ($\mathrm{d}I/\mathrm{d}V$) or its derivative ($\mathrm{d}^2I/\mathrm{d}V^2$) with respect to applied voltage. For example, in a normal-normal junction (N/N), a measurement of the conductance derivative yields information about the spectral weight of interactions, which are typically electron-phonon in nature.\cite{Jansen80, Yanson74}

New phenomena occur when the point-contact is between a normal metal and a superconductor (N/S).  For example, if there is no barrier between the two metals, the current through the junction increases by a factor of two below a certain energy $\Delta$. This energy is associated with the formation of Cooper Pairs in the superconductor. On the other hand, for a large barrier the current through the junction goes to zero for energies less than $\Delta$.

\begin{figure}
   \centering  
         \includegraphics[width=0.4\textwidth]{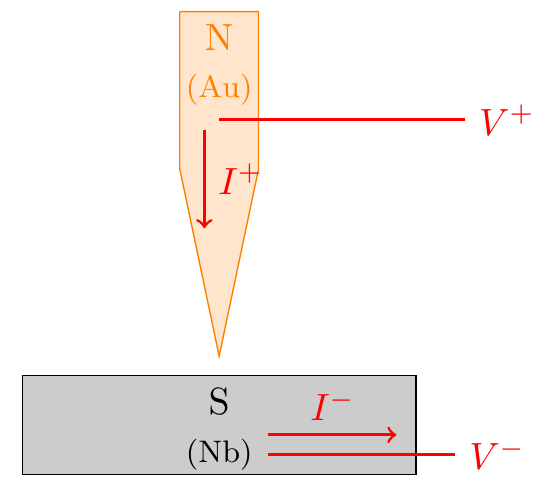}
   \caption{(Color online) Schematic of a point contact measurement where a normal metal tip touches a superconducting sample.  A differential conductance measurement is made by applying a stepped DC current with small AC modulation across the junction while the AC voltage is measured.}
   \label{LunaFig01}
\end{figure}

Point-contact spectroscopy of N/S junctions has been predominantly used to measure the energy gap and associated density of states. Using this information, lower bounds on the Fermi velocity and coherence length (the spatial size of a Cooper pair) of the superconductor can be obtained. Measurements on superconductors to obtain such information have been performed on materials such as niobium (Nb),\cite{btk, nbptex, laura} cuprates,\cite{deutscher}  $\mathrm{Sr}_2\mathrm{RuO}_4$,\cite{Laube00} $\mathrm{MgB}_2$,\cite{Naidyuk02} and heavy fermion materials.\cite{Park08} In this paper, we present an experiment, suitable for an advanced undergraduate laboratory, to study the N/S spectroscopic behavior of a point-contact between a gold (Au) tip and Nb film (see Fig.~\ref{LunaFig01}).
 
In order to understand how reflection and transmission occur at the interface between a normal metal and a superconductor, it is important to understand some basic facts about superconductivity.  At the transition temperature  of a superconductor $T_{\mathrm{c}}$, the resistance drops to zero and an energy gap $\Delta\!\sim\! 10^{-3}$ eV, opens up in the density of states.  As explained by Bardeen-Cooper-Schrieffer (BCS) theory,\cite{tinkham} this energy gap---also known as the pair-potential energy or binding energy---arises because it is thermodynamically more favorable for electrons of opposite spin to bind together into a spin-zero boson called a ``Cooper pair.''  Hence at $T=0$, when there are no thermal fluctuations, the electrons will exist only in Cooper pairs.

\begin{figure}
   \centering
        \includegraphics[width=0.4\textwidth]{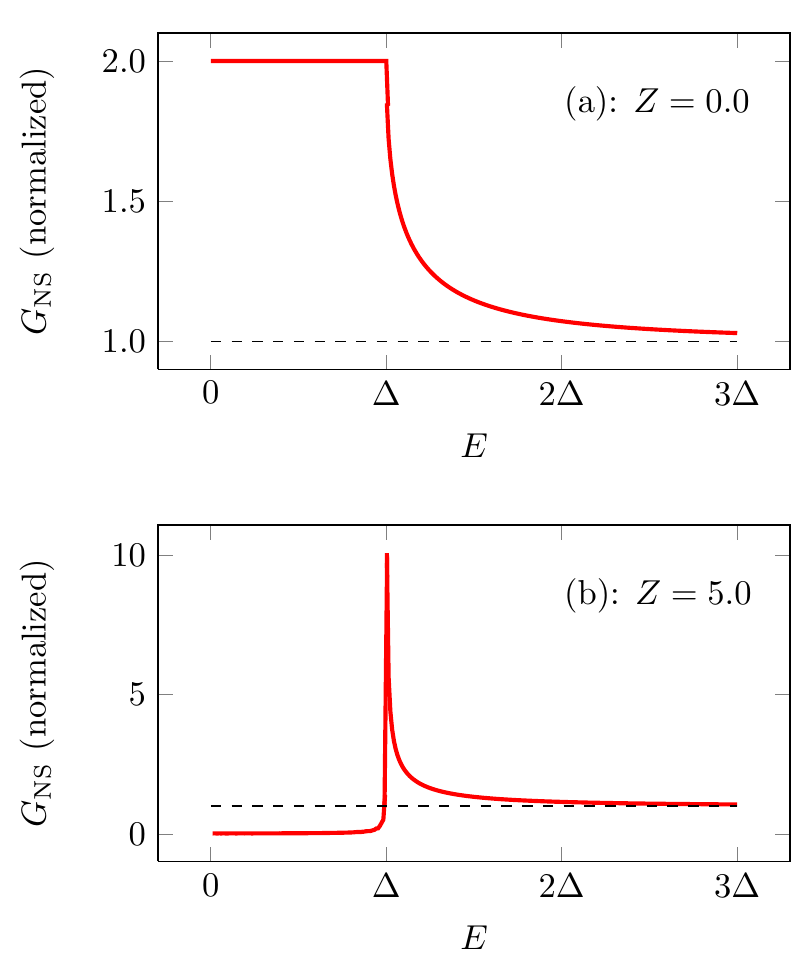}
   \caption{(Color online) Normalized differential conductance at positive bias voltage in the (a) Andreev and (b) tunneling regimes.  In Andreev reflection the differential conductance increases by a factor of two in comparison to the normal state above the gap energy $\Delta$.}
   \label{LunaFig02}
\end{figure}

In N/S point-contact transport measurements, three major processes occur at the interface: (1) transmission, (2) reflection, and (3) Andreev reflection. All can occur to varying degrees, although differences in the size of the junction and the tunneling barrier can alter the probabilities for any given event to occur.  The ``height" of the barrier at the N/S interface can be characterized by a dimensionless parameter $Z$; a low barrier that is easy to cross corresponds to $Z\approx 0$ while a high barrier that inhibits current is characterized by $Z>1$.   The resulting differential conductance $G_{\mathrm{NS}}$ for high and low barriers at $T=0$  is shown in Fig.~\ref{LunaFig02}.

Let us now return to a description of the three processes mentioned above. For energies above the gap ($E>\Delta$), Cooper pairs break and form single particle states in the superconductor. Such states are called ``quasiparticles'' because a particle and its influence on the local environment move together as a single particle-like entity.  In this energy regime, process (1) transmission occurs.  Here, single electrons are transmitted into the superconductor and manifests itself as a linear $I$-$V$ curve (constant differential conductance).

For energies below the gap ($E<\Delta$), the differential conductance depends on the constriction and tunneling barrier.  In the case of a large constriction where transport is not in the ballistic regime, process (2) reflection occurs independent of whether the tunneling barrier $Z$ is small or large.  In this case a single particle fermion cannot be transmitted into the superconductor which is composed of two-particle bosonic states.  The electrons from the normal metal therefore reflect from the interface of the superconductor and hence the current and $\mathrm{d}I/\mathrm{d}V$ are zero for this energy regime. Right at the gap energy $\Delta$, there is a discontinuity in the $I$-$V$ curve that gives rise to a peak in $\mathrm{d}I/\mathrm{d}V$.  The width of this peak is temperature dependent as broadening can occur due to thermal fluctuations that destroy the Cooper pairs.

\begin{figure}[bottom]
   \centering
\includegraphics[width=0.4\textwidth]{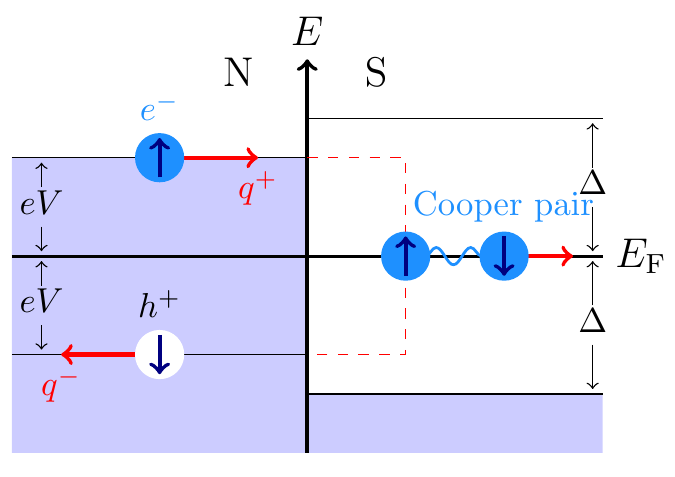}
   \caption{(Color online) A diagram showing Andreev reflection, where an electron in the normal metal with energy $E=E_\mathrm{F}+eV$ (and $eV<\Delta$) is incident upon a superconductor.  Note that in the body of the paper, $E_{\mathrm{F}}$ is referred to as $E=0$ because only excitations above this energy are considered in the theory.}
   \label{LunaFig03}
\end{figure}

If the constriction is small enough to be in the ballistic regime, then for a small tunneling barrier with $E<\Delta$,  process (3) Andreev reflection occurs and the differential conductance increases by a factor of two from the conductance above the gap (see Fig.~\ref{LunaFig02}a). Physically, this process can be characterized as shown in Fig.~\ref{LunaFig03}, where in order for a spin-up electron that is injected with energy $E<\Delta$ to be trasmitted, it must form a Cooper pair with an electron of opposite spin in the superconductor. Consequently, a hole---a mathematical construction to describe the absence of an electron---is reflected back with opposite spin and momentum of the incident electron. Hence if the incident electron carries a current $ev_{\mathrm{F}}$, where $v_{\mathrm{F}}$ is the Fermi velocity, then a current of $-ev_{\mathrm{F}}$ is reflected as a hole. The net result is a current of $2ev_{\mathrm{F}}$ in the normal metal and a current of $2ev_{\mathrm{F}}$ in the superconductor as a result of the Cooper pair.

In a real N/S contact, a tunneling barrier will exist and the resultant differential conductance measurement will be somewhere in-between the two extreme tunneling barrier cases of Fig. \ref{LunaFig02}.  However, in the experiment presented in this paper, by using a differential screw to gradually press an Au tip into superconducting Nb, the tunneling barrier $Z$ can be varied over a range of values.

This paper is organized as follows: In section \ref{BTK Model} we discuss the Blonder-Tinkham-Klapwijk (BTK) theory\cite{btk} and derive the differential conductance and other relevant physical quantities.   In section \ref{Experiment} we discuss our differential screw-dipping probe apparatus and the circuit used to obtain the conductance curve as a function of the bias voltage applied across the junction.  Finally, in section \ref{Results} we present the results of our own differential conductance measurements, including an estimate of the energy gap and lower bounds for the BCS coherence length and Fermi velocity of Nb.

\section{The BTK Model} \label{BTK Model}
   
\begin{figure}[bottom]
   \centering
   \includegraphics[width=0.4\textwidth]{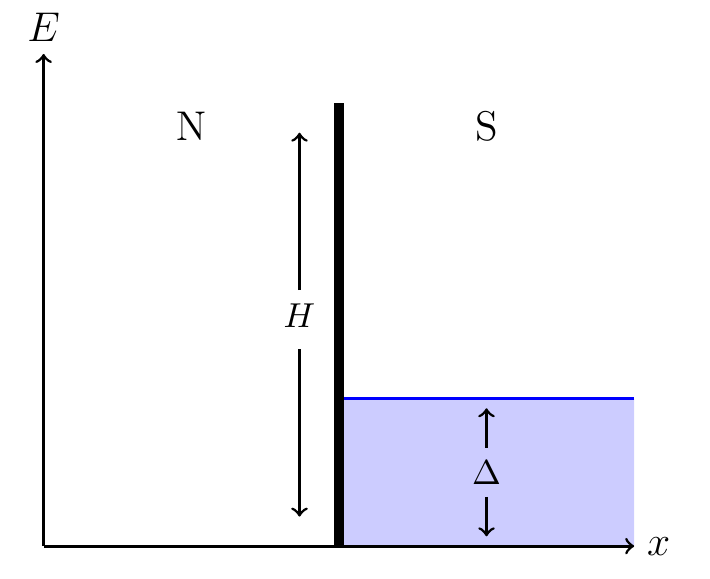}
   \caption{(Color online) Potential energy diagram for the BTK model.  There is a constant potential energy $\Delta$ in the superconductor and a barrier at the N/S interface modeled by a $\delta$-function of strength $H$.}
\label{LunaFig04}
   \end{figure}

The BTK theory used to model the point-contact junction considers the transmission, reflection, and Andreev reflection of a lone electron at the boundary between a normal metal and a superconductor.\cite{btk}  We take the potential to be a $\delta$-function barrier of amplitude $H$ at the interface, and a constant $\Delta$ inside the superconductor (see Fig.~\ref{LunaFig04}).  The incident wave is assumed to be a one-dimensional (1D) plane wave traveling from the normal metal into the superconductor.  A 1D model is sufficient for Nb as its gap is known to be spherically symmetric in momentum space.  We first calculate the probabilities for reflection and transmission of an electron at the N/S interface using familiar techniques learned in quantum mechanics.  Afterwards, we use the results to determine an expression for the macroscopic current that will be observed in our apparatus.

\subsection{Solution of the BTK Model}

In order to find the probabilities of the various processes, we solve the Bogoliubov-de Gennes (BdG) equations,\cite{btk} which are slightly modified versions of Schr\"odinger's equation.  If we denote the wave function by
\begin{equation}
\psi = \left(\begin{array}{c} f(x,t) \\ g(x,t) \end{array}\right),
\end{equation}
where $f(x,t)$ and $g(x,t)$ represent the electron and hole wave functions respectively, then the BdG equations are
\begin{align}
\mathrm{i}\left(\hbar\frac{\partial }{\partial t} +\Gamma\Theta\left(x\right)\right)\sigma_z \psi &= \left(-\frac{\hbar^2}{2m}\frac{\partial^2}{\partial x^2} - \mu(x)+V(x)\right)\sigma_z\psi \notag \\
&\; \; \;+ \Delta(x)\sigma_x\psi.
\end{align}
Here, $\sigma_x$ and $\sigma_z$ are Pauli spin matrices
\begin{align}
\sigma_x &= \left(\begin{array}{cc} 0 &\ 1 \\ 1 &\ 0 \end{array}\right), \\
\sigma_z &= \left(\begin{array}{cc} 1 &\ 0 \\ 0 &\ -1\end{array}\right),
\end{align}
$\Theta(x)$ is the step function
\begin{equation}
\Theta(x) = \left\lbrace \begin{array}{ll} 1 &\ x\ge 0 \\ 0 &\ x<0 \end{array}\right.,
\end{equation}
$V(x)$ is the potential energy,  $\mu(x)$ is the chemical potential, and $\Delta(x)$ represents the gap.  The rate at which quasiparticles decay to the BCS ground state inside the superconductor is represented by the parameter $\Gamma$; if $\tau$ is the lifetime of these particles, then \begin{equation}
\Gamma = \frac{\hbar}{\tau}.
\end{equation}

The effective Schr\"odinger-like equation for this system looks like a generalized version of the traditional problem of quantum tunneling through a $\delta$-function barrier with a few extra terms.  The parameter $\Gamma$ governs quasiparticle decay through scattering and/or recombination effects.\cite{Kaplan76,Dynes78}  In practice, $\Gamma$ serves as a phenomenological parameter that broadens the peak shown in Fig. \ref{LunaFig02}b. The chemical potential is the energy required to add a particle to the system; since both electrons and holes are fermions, this is equivalent to the Fermi energy of the system at $T=0$.     The presence of the Pauli matrices is a convenient way to represent electrons, holes, and their interactions into one equation.  Phenomenologically, the term $\Delta \sigma_x$ allows for a coupling between electrons and holes.

In the BTK model, we consider the case where $\mu(x)=\mu$ and $\Delta(x)=\Delta\Theta(x)$ for constant $\mu$ and $\Delta$, and $V(x)=H\delta(x)$.  To characterize the height of the potential barrier at the interface, it is convenient to define a dimensionless parameter 
\begin{equation}
Z\equiv \frac{H}{\hbar v_{\mathrm{F_S}}},
\end{equation}
where $v_{\mathrm{F_S}}$ is the Fermi velocity of the superconductor.  To solve for the steady-state plane-wave solutions at the N/S interface, we use trial wave functions of the form
\begin{align}
\psi_{\rm{i}} &= \left(\begin{array}{c} 1 \\ 0 \end{array}\right)\mathrm{e}^{\mathrm{i}q^+ x}\mathrm{e}^{-\mathrm{i}Et/\hbar},\\
\psi_{\rm{r}}&=\left[a\left(\begin{array}{c} 0 \\ 1 \end{array}\right)\mathrm{e}^{\mathrm{i}q^- x}+b\left(\begin{array}{c} 1 \\ 0 \end{array}\right)\mathrm{e}^{-\mathrm{i}q^+ x}\right]\mathrm{e}^{-\mathrm{i}Et/\hbar},\label{refl} \\
\psi_{\rm{t}}&=\left[c\left(\begin{array}{c} u_0 \\ v_0 \end{array}\right)\mathrm{e}^{\mathrm{i}k^+ x}+d\left(\begin{array}{c} v_0 \\ u_0 \end{array}\right)\mathrm{e}^{-\mathrm{i}k^-x}\right]\mathrm{e}^{-\mathrm{i}(E+\mathrm{i}\Gamma)t/\hbar}.
\label{trans}
\end{align}
These incident, reflected, and transmitted waves lead to a more complex version of a common undergraduate problem on transmittance through a $\delta$-function barrier.

There are several important points to make regarding the form of $\psi$.  Firstly, there are four distinct wave numbers $k^\pm$ and $q^\pm$.  The reason to distinguish between $k$'s and $q$'s is because the inclusion of $\Theta(x)$ alters the dispersion relation $\omega(k)$ inside of the superconductor. Also, the signs on the diagonal of $\sigma_z$ require that we distinguish between the $+$ and $-$ wave numbers.  Secondly, inside the superconductor the waves are no longer pure electrons and holes but instead a combination of both---they are {\em quasiparticles}.  Because the Hamiltonian is no longer a diagonal matrix in the electron-hole basis, its eigenvectors will have two non-zero components.  For any given energy there are two quasiparticle states and we denote the one with a positive (negative) group velocity by $k^+$ ($k^-$).    Finally, a reflected hole propagates with a wave number of $+q^-$ and not $-q^-$.   Because a hole is the absence of a particle, differentiating the dispersion relation gives a group velocity with the opposite sign.   One of the quasiparticles will propagate forward with a wave number of $-k^-$ for the exact same reason (see Fig. 11.7 of Ref. \onlinecite{tinkham} for a nice picture of this dispersion relation).

The boundary conditions for this problem are continuity of the wave function
\begin{equation}
\psi_{\mathrm{N}}(0) = \psi_{\mathrm{S}}(0),
\end{equation}
where $\psi_\mathrm{S} = \psi_{\mathrm{t}}$ and $\psi_\mathrm{N} = \psi_{\mathrm{i}} + \psi_{\mathrm{r}}$, and a discontinuity in the slope of the wave function across the $\delta$-function
\begin{equation}
-\frac{\hbar^2}{2m}\left(\frac{\mathrm{d}\psi_{\mathrm{S}}(0)}{\mathrm{d}x} - \frac{\mathrm{d}\psi_{\mathrm{N}}(0)}{\mathrm{d}x}\right) + H\psi_{\mathrm{S}}(0)=0.
\end{equation}
Using these conditions, we can derive the relations 
\begin{align}
\hbar q^\pm &= \sqrt{2m(\mu \pm E)}, \\ 
\hbar k^\pm &= \sqrt{2m\left(\mu \pm \sqrt{\left(E+\mathrm{i} \Gamma\right)^2-\Delta^2} \right)} ,
\end{align}
and
\begin{equation}
1-v_0^2 =u_0^2 =\frac{1}{2}\left(1+\sqrt{\frac{\left(\left(E+\mathrm{i} \Gamma\right)^2-\Delta^2\right)}{\left(E+\mathrm{i} \Gamma\right)^2}}\right).
\end{equation}
We can solve for $a$ and $b$ from Eq.~\eqref{refl} if we approximate $q^+=q^-=k^+=k^-=k_\mathrm{F}$ (or equivalently $\mu\gg E,\;\Delta,\;\Gamma$).  Because we are dealing with energies of only a few meV whereas a typical (normal) metal has a Fermi energy of 1--10 eV, this is a reasonable approximation for any normal metal (see Table 2.1 of Ref. \onlinecite{mermin}).  We then obtain
\begin{align}
a &=\frac{u_0v_0}{\gamma}, \label{a} \\
b &=-\frac{\left(u_0^2-v_0^2\right)\left(Z^2+\mathrm{i}Z\right)}{\gamma},\label{b}
\end{align}
where
\begin{equation}
\gamma =u_0^2+\left(u_0^2-v_0^2\right)Z^2.
\end{equation}
Using these results, we can calculate the probabilities of Andreev reflection $A(E)=aa^*$ and normal reflection $B(E)=bb^*$.  Lastly, because Nb and Au have different Fermi velocities, we must use an effective $Z_{\rm{eff}}$ that satisfies
\begin{equation}
{Z_{\rm{eff}}}^2 = Z^2 + \frac{(1-r)^2}{4r},
\label{Z}
\end{equation}
where $r = v_{\mathrm{F_N}}/v_{\mathrm{F_S}}$ is the ratio of the Fermi velocities in the normal and superconducting metals.\cite{zeffarticle}  By using $Z_{\rm{eff}}$ we are able to place a lower bound on the Fermi velocity of the superconductor.

\subsection{Differential Conductance}

We now derive the differential conductance of a N/S tunnel junction from first principles without Andreev reflections and show how it is related to the density of states of the superconductor.  We will then incorporate Andreev reflection using the BTK model, and show how to estimate the energy gap and place a lower bound on the Fermi velocity and BCS coherence length.

Ignoring Andreev reflections for the moment, the probability for a tunneling process is the product of the number of occupied initial states, the number of unoccupied final states, and the probability that the tunneling event occurs.  For tunneling between the normal metal to the superconductor, the number of occupied initial states is $N_\mathrm{N}(E)f(E)$ and the number of unoccupied final states is $N_{\mathrm{S}}(E+eV)[1-f(E+eV)]$.  Here, $N_{\mathrm{N}}(E)$ is the density of states of the normal metal, $N_{\mathrm{S}}(E)$ is the density of states of the superconductor, and $f(E)$ is the Fermi function.  For the reverse current, the joint probability for this process is $N_{\mathrm{N}}(E)f(E+eV)N_{\mathrm{S}}(E+eV)[1-f(E)]$.  Hence if we subtract the reverse current from the forward current, we obtain
\begin{equation}
I_{\mathrm{NS}} =\alpha|T|^2 \int_{-\infty}^\infty\! N_\mathrm{N}(E)N_\mathrm{S}(E+eV)[f(E)-f(E+eV)]\,\mathrm{d}E,
\label{Itransfer}
\end{equation}
where $\alpha$ is a constant that depends on the area of the junction and the Fermi velocity (see pp. 44-47 of Ref. \onlinecite{tinkham}).  $|T|^2$ is a phenomenological tunneling matrix element that depends on the type of insulator between the N/S junction and governs whether the event occurs.  The difference in Fermi functions $f(E)-f(E+eV)$ represents the probability of transmission, where $V$ is the applied voltage and $eV$ the resulting difference in the chemical potential across the junction.

 In the energy regime in which we are dealing, we can approximate $N_{\mathrm{N}}(E)=N_{\mathrm{N}}(0)$ as constant since $E\ll \mu$ has been previously assumed.  Using the current in Eq. (\ref{Itransfer}), the differential conductance is then
\begin{equation}
\frac{\mathrm{d}I_{\mathrm{NS}}}{\mathrm{d}V}\propto \int_{-\infty}^\infty N_{\mathrm{S}}(E)\left[-\frac{\partial f(E+eV)}{\partial (eV)}\right]\,\mathrm{d}E,
\end{equation}
where we have assumed that $N_{\mathrm{S}}(E)$ is slowly varying compared to the Fermi weighting factor.  Note that $-\partial f(E+eV)/\partial (eV)$  is a bell-shaped weighting function with a maxinum at $E=-eV$, a width of $\sim k_{\mathrm{B}}T$, and unit area.  This factor plays the same mathematical role as a Dirac delta function inside an integral when $k_{\mathrm{B}}T\rightarrow 0$.  At zero temperature, the differential conductance approaches the density of states of the superconductor
\begin{equation}
G_{\mathrm{NS}}\equiv \left. \frac{\mathrm{d}I_{\mathrm{NS}}}{\mathrm{d}V}\right|_{T=0}\propto N_S(e|V|),
\label{dIdV}
\end{equation}
which is equivalent to what is shown in Fig.~\ref{LunaFig02}b for large $Z$.

If we use the BTK formulation to incorporate Andreev reflections, the current that flows across the junction is
\begin{equation}
I_{\mathrm{NS}} \propto \int_{-\infty}^\infty (1+A(E)-B(E)) (f(E)-f(E+eV))  \mathrm{d}E.
\label{Ins}
\end{equation}
In this equation, $1+A(E)-B(E)$ represents that transmission coefficient for electrical current.  This coefficient incorporates the fact that when a hole is reflected, a positive charge moves in the opposite direction as the incident electron. Such a current is represented by $A(E)$ and will add to the incident current. Similarly, reflected electrons are represented by $B(E)$ and will reduce the current. Although the current obtained in Eq. (\ref{Ins}) using the BTK approach is similar to Eq. (\ref{Itransfer}), it is different in the fact that the normal metal and superconductor are not decoupled and the initial wave packet evolves continuously through the barrier (see section V of Ref. \onlinecite{btk}). However, the first approach is important from a pedagogical standpoint, as it is easy to interpret the relationship between the tunneling current and the density of states of the superconductor.

Using the current in Eq.~(\ref{Ins}), the differential conductance of a S/N junction at low temperature is approximately given by the transmission coefficient for electrical current
\begin{equation}
G_{\mathrm{NS}} = \left.\frac{\mathrm{d}I_{\rm{NS}}}{\mathrm{d}V}\right|_{T=0} \propto (1+A(eV)-B(eV)).  \label{Gns}
\end{equation}  This conductance is also related to the density of states of the superconductor, but the relationship is harder to discern because the density of states is encoded in the coherence factors $N_{\mathrm{S}}(E)=(u_0^2-v_0^2)^{-1}$.  A benefit to this BTK formulation is that the barrier height is not assumed to be large so that Andreev reflection processes come into consideration, whereas it is absent in the formulation given by the current in Eq.~(\ref{Itransfer}).  Therefore, Eq.~(\ref{Gns}) provides a spectrum of conductances from a low-$Z$ barrier (Fig.~\ref{LunaFig02}a) to a high-$Z$ barrier (Fig.~\ref{LunaFig02}b).

By measuring the differential conductance of a Nb/Au junction and fitting the experimental data to the BTK model using the derivative of Eq.~(\ref{Ins}), we can obtain values for $Z_{\rm{eff}}$, $\Delta$, and $\Gamma$.  The latter two are of fundamental interest.  In addition, we can use the value for $Z_{\rm{eff}}$ to obtain a lower bound on the Fermi velocity in the superconductor.  Letting $Z=0$ and using Eq.~(\ref{Z}) to obtain a lower bound on the Fermi velocity ratio $r$, we find the BCS coherence length $\xi$ of the quasiparticle waves in Nb\cite{deutscher} to be
\begin{equation}
\xi = \frac{\hbar v_{\mathrm{F_S}}}{\pi \Delta}.
\end{equation}

It is important to note that the BTK model assumes ballistic transport, which implies that the size of the contact $a$ is less than the electron mean free path $l$ and the coherence length $\xi$.  If $a\ll l$, the contact is ballistic which prevents heating effects at large current densities at voltages on the order of the gap energy.  The resistance in this regime, known as the Sharvin resistance, $R_0 = \rho l / 4a^2$, is due to the radius of the constriction from the normal metal to the superconductor.  Furthermore, the size of the contact must be much less than the coherence length ($a\ll \xi$) to suppress the proximity effect, where the tip of the gold begins to weakly superconduct and the niobium becomes an imperfect superconductor.  Additionally, this condition prevents the destruction of superconductivity near the gap where the carrier velocity approaches the depairing value ($\Delta / p_{\mathrm{F}}$), where $ p_{\mathrm{F}}$ is the Fermi momentum.\cite{deutscher}

In a typical low temperature superconductor, the coherence length is on the order of $1000$ \r{A} and the contact in the Sharvin regime is around $a \approx 100$ \r{A}.  However, typical tips, such as a gold wire cut by a razor blade, or in our case an electrochemically etched gold tip, have a radius on the order of $0.1-1$ $\mu$m.  A tip this blunt should have a contact resistance around $ 1-10\; \rm{m}\Omega$. In practice, the contact resistance is often found to be around $10-100 \;\Omega$, corresponding to a much smaller contact area.  Various explanations have been proposed for this observation, but there is no consensus as to why the effective contact size is so good in practice. Nevertheless, data can be found that fits well with BTK theory.\cite{deutscher}

\subsection{Fitting Experimental Data to the BTK Model}

We now explain how to fit the theoretical model to obtain the parameters $\Delta$,  $\Gamma$, and $Z_{\mathrm{eff}}$. In Eq. (\ref{Gns}), the constant of proportionality is unknown, so in order to compare the experimental and theoretical results, the data is normalized by a value of the differential conductance above the gap.  It is not uncommon for there to exist a background slope due to thermal voltages or other external factors.  To try to minimize this effect, we applied the following normalization convention \begin{equation*}
\frac{1}{2}\left(\lim_{V\rightarrow -\infty} G_{\mathrm{NS}}(eV) +\lim_{V\rightarrow \infty} G_{\mathrm{NS}}(eV) \right)=1,
\end{equation*}
which is shown as a dashed line in Fig.~\ref{LunaFig02}.

In order to perform a non-linear least squares fit of the data to the theoretical model, an array of theoretical conductances given measured voltages needs to be constructed.  In our case, the temperatures are significantly below $T_{\mathrm{c}}$, so we can approximate the differential conductance as in Eq. (\ref{Gns}). It is convenient to use MATLAB or some other program to expand the quantities in this equation, particularly since $a$ and $b$ are complex. The measured voltage values are then taken as the input for the theoretical differential conductance $G_{\mathrm{theor}}(eV)$ in Eq. (\ref{Gns}).

For cases where the temperature is closer to $T_{\mathrm{c}}$, we can perform a Riemann sum and cut off the limits of integration because the integrand is non-zero for a small range localized about $eV$.  Both methods were used for our data analysis in order to confirm that allowing $T=0$ did not alter the values of the parameters.

Together, the values of $\Delta$, $\Gamma$, and $Z_{\mathrm{eff}}$ parameterize the differential conductance of Eq.~(\ref{Gns}).   An additional parameter was included to account for any voltage offest.  A nonlinear least squares curve fit with MATLAB was used to find the values of these parameters where the theoretical and experimental curves most closely matched.

\section{Experimental Setup} \label{Experiment}
The Nb/Au junction is formed by having a differential screw that allows an electrochemically etched Au tip to gradually touch down onto a Nb film.  The Nb film was made in an evaporator that was baked overnight to minimize the amount of water in the chamber, as excess oxygen during deposition decreases the transition temperature.  A silicon substrate was ion milled for a few seconds in order to remove excess water or contaminants.  Then 100 nm of Nb was evaporated onto the substrate at a rate of 2 \r{A}/s at a base pressure of $\sim 10^{-7}$ Torr.  We independently verified a transition temperature $T_\mathrm{c}\approx 7$ K for the sample using a four-point measurement of resistance vs. temperature.

The sample was adhered with Lakeshore VGE-7031 insulating varnish to a chip holder made from a small PC board affixed with Loctite Hysol 1C epoxy to a copper beryllium plate.  Wire bonds were used to make electrical contact between the PC board and the sample.  To measure the differential conductance through the junction, two wires from the sample are necessary for a voltage and a current lead. However, four contacts were placed on the sample to perform a four point measurement of the resistance to verify that the sample became superconducting when placed in liquid helium.  We then varnished the copper-beryllium plate to the base of our apparatus.

\begin{figure}[bottom]
\centering
\includegraphics[width=0.4\textwidth]{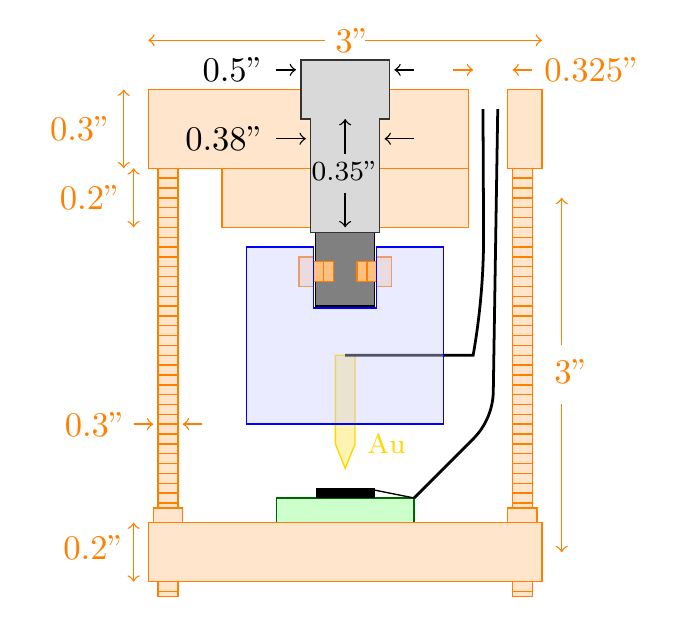}
\caption{(Color online) A cross-section of the bottom end of the cylindrical probe (not to scale).   All disks, rods, and holes are cylindrical and all parts are fixed except for the inner rod (dark gray).  The outer rod (light gray) has a smaller diameter at the end in order to reduce horizontal motion of the inner rod.  The Nb sample is on a removable plate that is varnished to the bottom disk.  Wires from both the Nb and Au are threaded up the probe through a hole cut out in the upper ring; each wire drawn corresponds to a pair of wires threaded around each other.}
\label{LunaFig05}
\end{figure}

The gold tip was electrochemically etched\cite{Libioulle95, Ren04, Lindau74, Baykul00, Nam95} starting from a 0.5 mm gold wire of 99.9985\% purity purchased from Alpha Aesar (product number 10966).   The tip acted as an anode while a platinum foil acted as a cathode in a diluted $40\%$ HCl solution that was heated to $50^{\circ}$C.  A function generator produced a 10 kHz square-wave signal with 5 V peak-to-peak and 2.5 V offset in order to act as the voltage source for the electrochemical reaction.  Immediately after etching, we rinsed the tip with hot deionized water to remove salts and then dipped it into acetone for further cleaning.  We then placed the gold tip into an insulating acrylic plastic holder attached to the apparatus (Fig. \ref{LunaFig05}).  While the plastic holder functioned well for our purposes, it did crack slightly.  For future work we recommend using macor or polycarbonate, both of which are insulating machinable materials that have better thermal contraction properties at cryogenic temperatures.

\begin{figure}[bottom]
\centering
\includegraphics[width=0.4\textwidth]{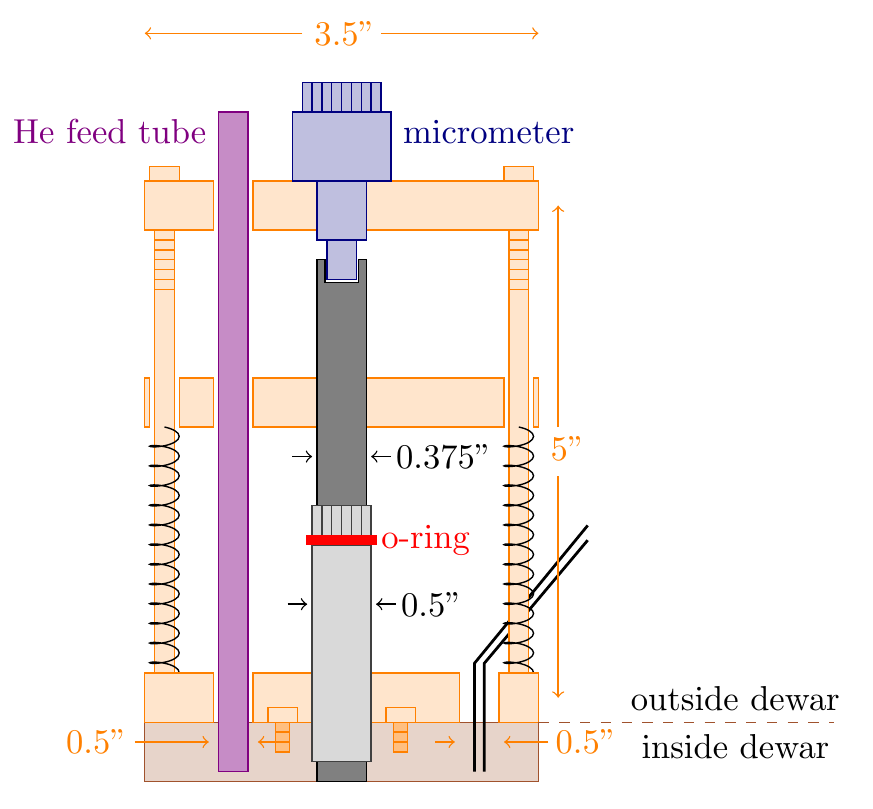}
\caption{(Color online) A cross-section of the top part of the probe (not to scale).  All holes and disks are cylindrical, with radii and heights specified where appropriate.   The inner most component of the micrometer is attached to the inner rod (light gray), which is free to move vertically.  The inner rod is attached to the central plate which is also free to move.  Springs attached to the three support rods help to dampen noise.  A nut with an O-ring seal, used in ultra-torr vacuum fittings, is used to loosely hold the inner rod in place with respect to the outer rod (dark gray) and aids in holding vacuum while measurements are taken. The wires are epoxied into a removable brass tube which is then inserted into a permanent tube and sealed with a nut and an O-ring. When pumping on the liquid helium, a rubber stopper is inserted into the hole used for transferring helium.  The bottom plate is screwed into the top of the dewar.}
\label{LunaFig06}
\end{figure}

A model DM-13 differential micrometer purchased from Newport controlled the vertical movement of the tip. It was mounted at the top of the probe outside the glass dewar cryostat and on top of springs that mitigated vibrations (Fig. \ref{LunaFig06}).  An O-ring seal held the dipping probe in place, but the force supplied by the differential screw  was sufficient to overcome this grip and allow us to push the probe further toward the sample.  The O-ring also provided a good enough vacuum to prevent the escape of helium gas from the bath.  In order to minimize unwanted horizontal movement of the tip, we welded an outer rod to the top plate to act as a fitted guide for the inner rod driven by the differential screw.  There was a guided bearing at the base of the outer rod that minimized the horizontal motion of the inner rod (see Fig. \ref{LunaFig05}).  Stainless steel was used to construct the inner and outer rods.  See supplementary material  for more detail on the schematics of this apparatus and the cost of the electronics.\cite{supplement}

The differential conductance was obtained for different input currents by using an AC modulating technique (p. 46 of Ref. \onlinecite{spectbook}).  A DC current $I_{\mathrm{DC}}$ and an AC current $I_{\mathrm{AC}}\cos(\omega t)$ with $I_{\mathrm{AC}}\ll I_{\mathrm{DC}}$ was applied across the junction and the voltage was measured.  Expanding the voltage in a Taylor series
\begin{align}
V(I_{\mathrm{DC}}+&I_{\mathrm{AC}}\cos(\omega t)) \approx V(I_{\mathrm{DC}}) \notag \\ &+ \frac{\mathrm{d}V(I_{\rm{DC}})}{\mathrm{d}I} I_{\mathrm{AC}}\cos(\omega t) + \cdots ,
\end{align}
we see that a measurement of the AC voltage determines $\mathrm{d}V/\mathrm{d}I$, the inverse of the differential conductance.

\begin{figure}[top]
   \centering
  \includegraphics[width=0.4\textwidth]{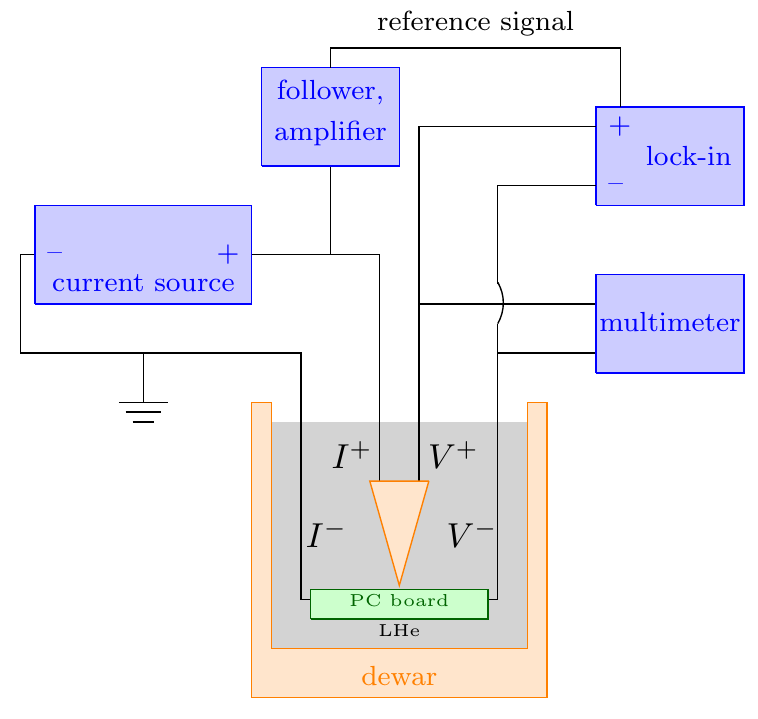}
   \caption{(Color online) A block diagram of the electronics.}
   \label{LunaFig07}
\end{figure}
Figure~\ref{LunaFig07} depicts a block diagram of the electronics.  A Keithley 6221 AC/DC current source provided the AC and DC currents across the junction.  We used a lock-in amplifier across the junction to measure the AC voltage and a digital multimeter to measure the DC voltage.  For contact resistances from 2--20 $\Omega$, an AC current of around 50 $\mu$A produced an AC voltage approximately an order of magnitude less than the gap for Nb ($\sim 1 $ meV).  In general the AC voltage used should be about an order of magnitude less than the width of the peak at the gap energy in order to resolve the signal.  If the AC voltage were larger than the peak width, then the differential conductance at the gap energy would be in essence an average of measurements not localized around the peak height, leading to an underestimate of its true value. We used a frequency of 1 kHz for the AC current and the DC current was stepped from negative to positive values and back so as to observe any heating effects or other anomalies.  For most measurements $|I_{\mathrm{DC}}| \le 600$ $\mu$A.

When the gold tip approaches the sample, a resistor in parallel with the source will prevent overloading the current source.  This resistor should be a few orders of magnitude greater than the expected resistance of the N/S junction.  Once the tip touches down onto the sample, the resistance will drop significantly, indicating that contact is made. At this point, the shunt resistor is removed from the circuit.  In addition to the electrical signal that is observed when contact is made, the glass dewar also has a slit to visually observe when the tip comes in contact with the sample.  The differential screw allows us to increase and decrease the pressure of the tip on the sample, thus changing the contact resistance for various measurements. 

While it is possible to have the Au tip in contact with the Nb film at room temperature, this can cause thermal contractions that damage the sharp tip and increase the contact area beyond that required to observe Andreev reflection.  Therefore, the tip and sample are separated at the start of the experiment.

\section{Results} \label{Results}

\begin{figure}[top]
\centering
\includegraphics[width=0.4\textwidth]{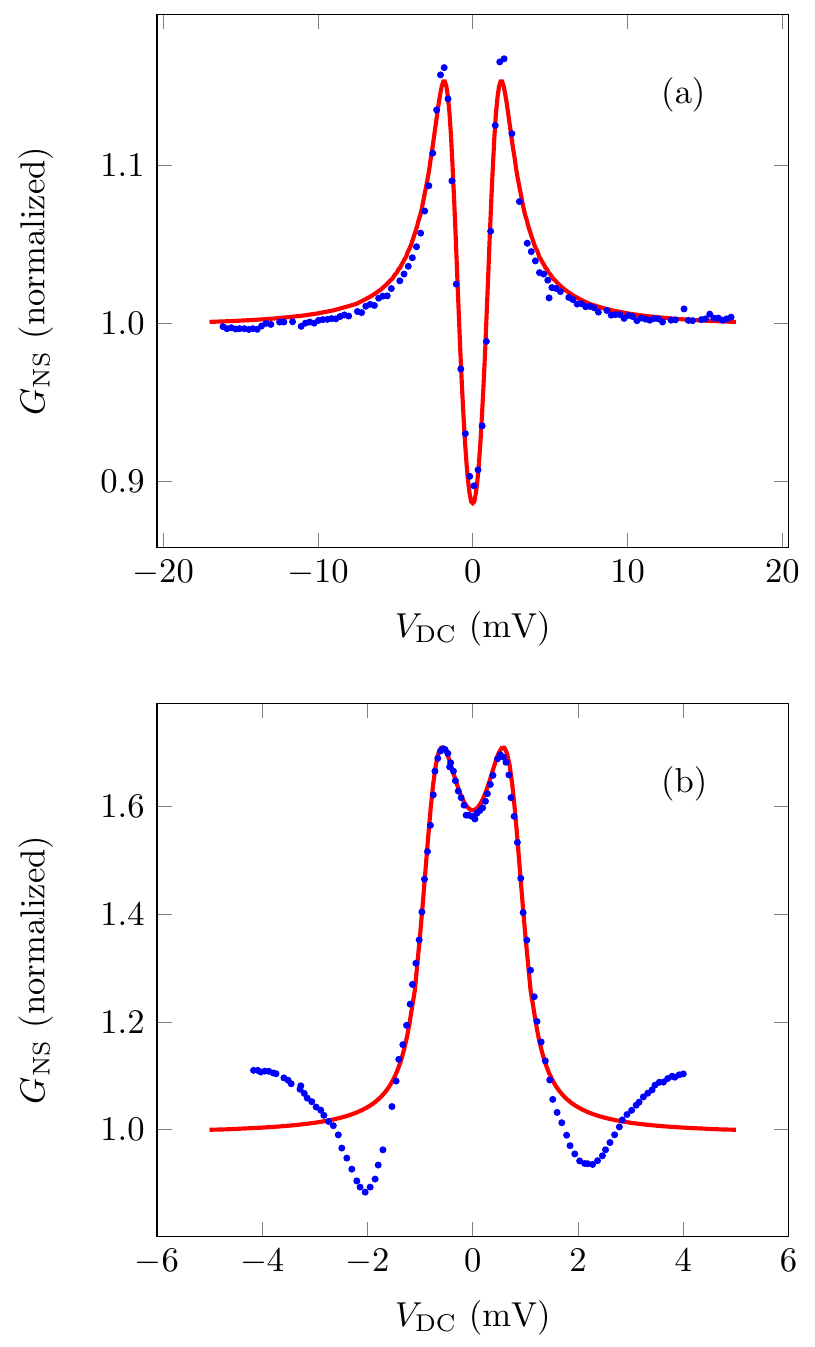}
\caption{(Color online) Normalized conductance with dots representing experimental data and solid curves representing a fit to BTK theory. (a) A sweep from negative to positive bias voltage with $V_{\rm{offset}}=0.33$ mV at 10-$\Omega$ contact resistance.  (b) A sweep from positive to negative bias voltage at a higher tip pressure, with $V_{\rm{offset}}=-0.11$ mV at 3-$\Omega$ contact resistance.}
\label{LunaFig08}
\end{figure}
Most of the our data is similar in character to the measurement shown in Fig.~\ref{LunaFig08}a (for contact resistances around $10$ $\Omega$) and Fig.~\ref{LunaFig08}b (for contact resistances around $3$~$\Omega$). These measurements were performed at a temperature of $T\approx 1.5$ K.  Table~\ref{LunaI} summarizes the fit parameters $Z_{\mathrm{eff}}$, $\Delta$ and $\Gamma$, which were found using a MATLAB least squares fit to Eq.~(\ref{Gns}).  For comparison, the fit parameters from Park et. al. are also provided.\cite{laura}

\begin{table}[]
\centering
\begin{tabular}{|c|c|c|c|}\hline
&\ 10 $\Omega$ &\ 3 $\Omega$  &\ Park et. al. \\ \hline
$Z_{\mathrm{eff}}$ &\ $0.82\pm 0.04$ &\ $0.29\pm 0.03$ &\ 0.453 \\ 
$\Delta$ (meV) &\ $1.53\pm0.10$ &\ $0.80\pm0.01$ &\ 1.25 \\
$\Gamma$ (meV) &\ $0.78\pm0.10$ &\ $0.02\pm 0.01$ &\ 0.55 \\\hline
\end{tabular}

\caption{Comparison of $Z_{\mathrm{eff}}$, $\Delta$ and $\Gamma$ as determined experimentally with both a 10 $\Omega$ and 3 $\Omega$ contact resistance.}
\label{LunaI}
\end{table}
For the data shown in Fig.~\ref{LunaFig08}a, the current was swept from negative to positive bias voltage and a DC offset of $0.33$ mV was used to shift the original data to be centered around zero voltage.  The offset used in fitting data for the sweep back from positive to negative bias was $\approx-0.11$ mV.  The reason for this offset is unknown, but could be due to an offset in our electronics or the presence of a time constant. From $Z_{\mathrm{eff}}$, we note that the differential conductance is neither fully in the Andreev reflection regime nor fully in the tunneling regime, but instead is a combination of the two.  This measurement was made after decreasing the tip pressure to increase the contact resistance.

Given our effective barrier strength $Z_{\rm{eff}}$ and the Fermi velocity of Au ($v_{\rm{F_N}} = 1.4\times 10^6$ m/s),\cite{deutscher} we estimate a lower bound on the Fermi velocity of Nb to be $3.1\times 10^5$ m/s and a BCS coherence length of 43 nm.  For comparison, an experiment that measured the superconducting upper critical field $H_{\mathrm{c2}}$ in bulk polycrystals found the Fermi velocity of Nb to be $v_{\rm{F_S}} = 5.7 \times 10^5$ m/s, consistent with our results.\cite{vfnb}  Additionally, Park et. al. performed point contact spectroscopy with a gold tip on a 210-nm film of Nb with $T_{\mathrm{c}} = 9.22$ K and obtained similar results after fitting their data to BTK theory.\cite{laura}  The pertinent results are listed in Tables~\ref{LunaI} and \ref{LunaII} for comparison.

\begin{table}[]
\centering
\begin{tabular}{|c|c|c|}\hline
&\ 10 $\Omega$ &\ Park et. al. \\\hline
$T_{\mathrm{c}}$ (K) &\ 7 &\ 9.22 \\
$v_{\mathrm{F_S}}$ (m/s) &\ $\ge 3.1\times 10^5$ &\ $\ge 5.8\times 10^5$ \\
$\xi$ (nm) &\ 43 &\ 97 \\\hline
\end{tabular}
\caption{Comparison of estimated values of $v_{\mathrm{F_S}}$ and $\xi$ for Nb.}
\label{LunaII}
\end{table}

As shown in Table~\ref{LunaII}, our measured coherence length is noticeably smaller than that reported by Park et. al.  Because of the impurities in our Nb thin film that reduce $T_{\rm{c}}$ to 7 K, we expect a smaller coherence length than for a pure Nb film with $T_{\rm{c}}$ of 9.2 K.  Other experiments have found the coherence length for bulk Nb to be around 40 nm.\cite{nbxi2} Hence our measurements are in reasonable agreement with the literature.

After pressing the tip further into the sample to reduce the contact resistance to 3 $\Omega$, we obtain significantly reduced values for $\Delta$ and $\Gamma$ (see Fig.~\ref{LunaFig08}b and Table~\ref{LunaI}) and a dip in the differential conductance at around  $\pm 2$ meV.  One possible explanation for the reduced gap is the proximity effect, in which $\Delta(x)$ departs from simple step function behavior at the interface of  S/N, and the end of the probe tip becomes weakly superconductivity.  The proximity effect is known to decrease the observed value for $\Delta$.\cite{proximityandreev, proximity, proximity2, laura}  Furthermore, the dip in the conductance was seen previously in a Nb/Cu point-contact spectroscopy experiment by Blonder and Tinkham who found that the tip had curled back on itself.  Blonder and Tinkham argue that this wider contact area allows heating to play a significant role.\cite{zeffarticle}  At this point, the initial assumptions of the BTK model---that $a\ll \xi$, $\ell$ and that the electrons are ballistic---begin to break down.\cite{deutscher, Sheet04}  We also observed a curled tip in our experiment.

\section{Conclusion}
We performed point-contact spectroscopy to measure the differential conductance as a function of bias-voltage through a Au/Nb junction. We then estimated the superconducting gap, quasiparticle lifetime, and placed a lower bound on the Fermi velocity and BCS coherence length in Nb by fitting the experimental data to theoretical curves predicted by the BTK model.

One of the largest drawbacks to this setup is the amount of time required to cool down a glass dewar cryostat before data can be acquired.  In addition, we only obtained approximately ten reasonable sets of data before a dip in the differential conductance appeared around $\pm 2$~mV.   As discussed above, this signifies a curled tip with large contact area that negates the underlying assumptions of the BTK model.  Unfortunately, at this point a new tip cannot be easily replaced mid-experiment because the cryostat must be warmed up to remove the tip. Furthermore, there were concerns about the possibility of the probe freezing due to thermal contraction of the cylindrical rods.  Such contraction would trap helium between the rods and render the probe immovable. Although such concerns were never realized in practice, this possibility could be eliminated by cutting a small slit in the outer rod to allow helium to escape.

This experiment easily extends to measure other superconducting materials or different types of junctions, such as a superconductor-superconductor junction.  Plus, the introduction of a magnetic field into the apparatus would allow for measurement of tunneling processes from a normal metal to a ferromagnet.  One improvement would be to measure incremental changes on the differential screw to provide quantitative data of the pressure between the tip and the sample.

The present experiment can be simplified by using a block of Nb rather than a thin film.  Films are easier to insulate and were readily available to us, but a Nb block can be placed on a glass slide and wires attached with silver epoxy. A superconducting sample can also be purchased from industry.  If one obtained a high-$T_{\mathrm{c}}$ sample (e.g. YBa$_2$Cu$_3$O$_{7-\delta}$), a simple setup can be constructed to immerse the sample into a liquid nitrogen container.  If liquid helium is required, one can reconfigure the apparatus described previously to fit inside of a storage dewar.  Because the height at which the dipping probe is inserted into the storage dewar stabilizes to a fixed temperature, such a setup would allow measurement of the differential conductance as a function of temperature.  Hence the transition of the differential conductance when the sample is above $T_{\mathrm{c}}$ to below $T_{\mathrm{c}}$ can be observed.

\section*{Acknowledgments}
We would like to thank Karlheinz Merkle and the Stanford Physics Machine Shop for machining much of our apparatus, Weigang Wang for his electronics expertise, Robert Hammond for making our Nb thin film, Rick Pam for maintaining the laboratory,  and David Goldhaber-Gordon, Malcolm Beasley, and Alexander Fetter for helpful discussions.










\begin{thebibliography}{99}
\bibitem{supplement}   [URL will be inserted by AIP]
\end{thebibliography}
\end{document}